\documentstyle[psfig,proceedings]{crckapb}
\makeatletter
\renewcommand{\subsection}{\@startsection
 {subsection}{2}{\z@}{.2ex plus .1ex minus .1ex}{-4pt}{\normalsize\it}}
\renewcommand{\fps@figure}{thbp}
\renewcommand{\fps@table}{thbp}

\makeatother
\def\floatwidth{0.7\textwidth}

\def\point#1{$\bigcirc\hspace*{-8pt}\mbox{\footnotesize #1}\;$}
\def\mag{\nobreak\mbox{$^{\rm m}$}\!\!\!\!.\;}

\def\kms{\nobreak\mbox{$\;$km\,s$^{-1}$}}

\begin{opening}
\title{The Linearity of the Cosmic Expansion Field and
       the Value of the Hubble Constant}

\author{G. A. Tammann}
\institute{Astronomisches Institut der Universit\"at Basel, \\
       Venusstr.~7, CH-4102 Binningen, Switzerland}

\date{}
\end{opening}

\runningauthor{G. A. Tammann}
\runningtitle{The Cosmic Expansion Field and H$_0$}

\begin{document}
\begin{abstract}
A linearity test shows $H_0$ to decrease by 7\% out to
$18\,000\kms$. The value at $10\,000\kms$ is a good approximation to
the mean value of $H_0$ over very large scales. The construction of
the extragalactic distance scale is discussed. Field galaxies, cluster
distances relative to Virgo, and blue supernovae of type Ia yield
$H_{0}$\,(cosmic) with increasing weight; they give consistently
$H_{0}=57\pm7$ (external error). This value is supported by purely
physical distance determinations (SZ~effect, gravitational lenses,
MWB~fluctuations). Arguments for $H_{0}>70$ are discussed and shown to
be flawed.
\end{abstract}

\section{Introduction}
\label{sec:1}
The calibration of the cosmic expansion rate $H_0$ consists of
two steps. The first step is an investigation of the cosmic
expansion field. How {\em linear\/} is the expansion? How large
are {\em systematic\/} deviations from linearity in function of
distance? What is the {\em scatter\/} of individual objects due
to peculiar motions about the mean expansion?
Only after these questions are solved can the second step be
tackled, i.\,e. the calibration of the expansion rate in absolute
terms.

   The procedural difference between the two steps is that only
redshifts and {\em relative\/} distances are needed for an
investigation of the characteristics of the expansion field, while
the calibration of the present large-scale expansion rate $H_0$
requires in addition the {\em true\/} distance of at least one
object which demonstratably partakes of the mean expansion.

   Much confusion about the expansion rate has arisen from
equating the velocity-distance ratio of a subjectively chosen
object with $H_0$. The determination of $H_0$ from the Virgo,
Fornax, or Coma clusters, for instance, is meaningful only if it
is demonstrated that they reflect at their moderate distances
the mean cosmic expansion. The long-standing problem of correcting the
observed mean velocity of the Virgo cluster into the frame of the
cosmic expansion field has become a classic
(cf. Section~3.2). The Fornax cluster with a velocity
of $v\approx1200\kms$ cannot be used for the determination of
$H_0$, even if a useful distance was known for it, because its
unknown peculiar velocity may well be as high as 20\% of its
observed velocity. And the Coma cluster at $v\approx7000\kms$,
which is sometimes used for the determination of $H_0$, may still
have a peculiar velocity component of 10\%, as the peculiar velocity
of $630\kms$ with respect to the MWB of one other supercluster, i.\,e.
the Local Supercluster, would suggest.

   The present paper outlines this two-step procedure. In
Section~2 the available data are used to map the expansion
rate in function of distance well beyond $30\,000\kms$, i.\,e.
out to distances where the truly cosmic character of $H_0$
cannot be questioned. Section~3 gives a summary of the
various methods
of determining distances of field galaxies, of the Virgo cluster,
and --- most decisively for $H_0$ --- of distant blue SNe\,Ia.
Methods leading seemingly to $H_{0}>70$ are
critically discussed in Section~4. A brief outlook is
given in Section~5.

\section{The Cosmic Expansion Field}
\label{sec:2}
\citeauthor{Hubble:29} (1929) in his discovery paper plotted recession
velocities versus {\em linear\/} distances to infer the expansion of
the Universe. All subsequent papers on the subject have used instead
a plot of $\log cz$ versus $\log ({\rm distance})$, the advantage
being that only distance {\em ratios\/} are needed.
The resulting diagrams have become known as Hubble diagrams.
As a measure of $\log ({\rm distance})$ the apparent magnitude of
standard candles, the apparent diameters of standard rods,
or relative distance moduli $\Delta(m-M)$ can be used.
Using the Hubble diagrams of different objects the overall
linearity of the cosmic expansion field has been proven
without doubt.

   In the case of linear expansion the regression line of the
Hubble diagram has slope 0.2 if apparent magnitudes of standard
candles or relative distance moduli are used. Their Hubble
diagrams allow therefore three additional tests:

\noindent
(1) If a specific data set yields a slope different from 0.2
it is an unfailing indication that the distance scale is
incorrect. (Uncorrected selection effects, i.\,e. Malmquist
bias, always yield too steep slopes and a spurious increase
of $H_0$ with distance).

\noindent
(2) The scatter about the regression line is due to a combination
of the effects of peculiar motions and errors of the relative
distances. If the latter are under control one can determine the
mean relative size $\Delta v/v_{\rm c}$ of the peculiar motions,
where $v_{\rm c}$ is the cosmic velocity required by the mean
Hubble line.

\noindent
(3) If the relative peculiar motions $\Delta v/v_{\rm c}$ are
plotted against the distance $r$ (or in sufficient approximation
against the recession velocity $v$) one can test for
{\em local\/} deviations from perfect Hubble flow.
If $H_{i}=(v_{\rm c} + \Delta v)/{r}$ is the perturbed
value of the Hubble ratio of the $i$-th object at distance
$r$, and $H_{0}= <\!\!H_{i}\!\!> ={v_{\rm c}}/{r}$ the true Hubble
constant, than ${\Delta H}/{H_0}=({H_{i}-H_{0}})/{H_0}=
{\Delta v}/{v_{\rm c}}$. This test of relative variation of $H_0$
is performed in the following.

   Three independent data sets are used for the test, viz.
the Hubble diagram of first-ranked cluster galaxies
\cite{Sandage:etal:76}, the Hubble diagram of
blue SNe\,Ia (\citeauthor{Saha:etal:97}, 1997; slightly updated by
\citeauthor{Parodi:Tammann:98}, 1998)
and the Hubble diagram with 31 relative cluster distances
\cite{Federspiel:etal:98}.
In each Hubble diagram the residuals ${\Delta v}/{v_{\rm c}}$
are read and combined within $5000\kms$ bins.
Sliding means in $2500\kms$ steps are plotted in Fig.~\ref{fig:1}
against redshift.
\def\floatwidth{0.65\textwidth}
\begin{figure}[t]
\centerline{\psfig{file=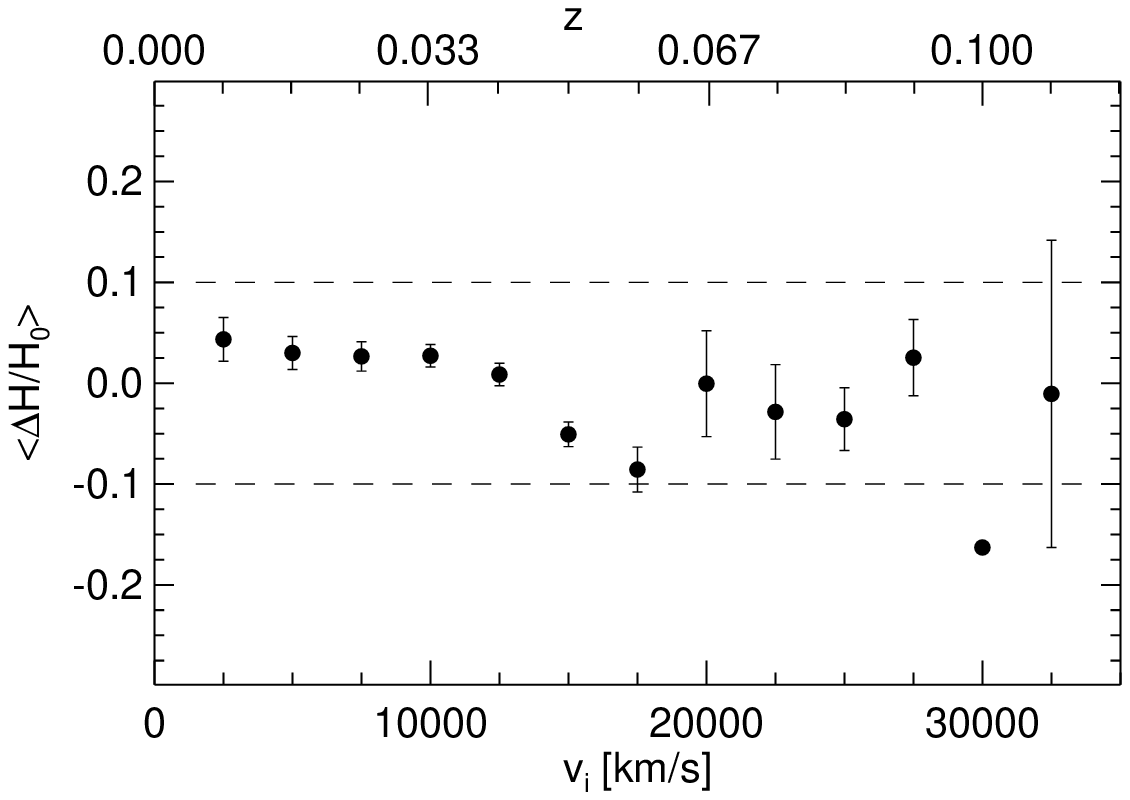,width=\floatwidth}}
\centerline{\psfig{file=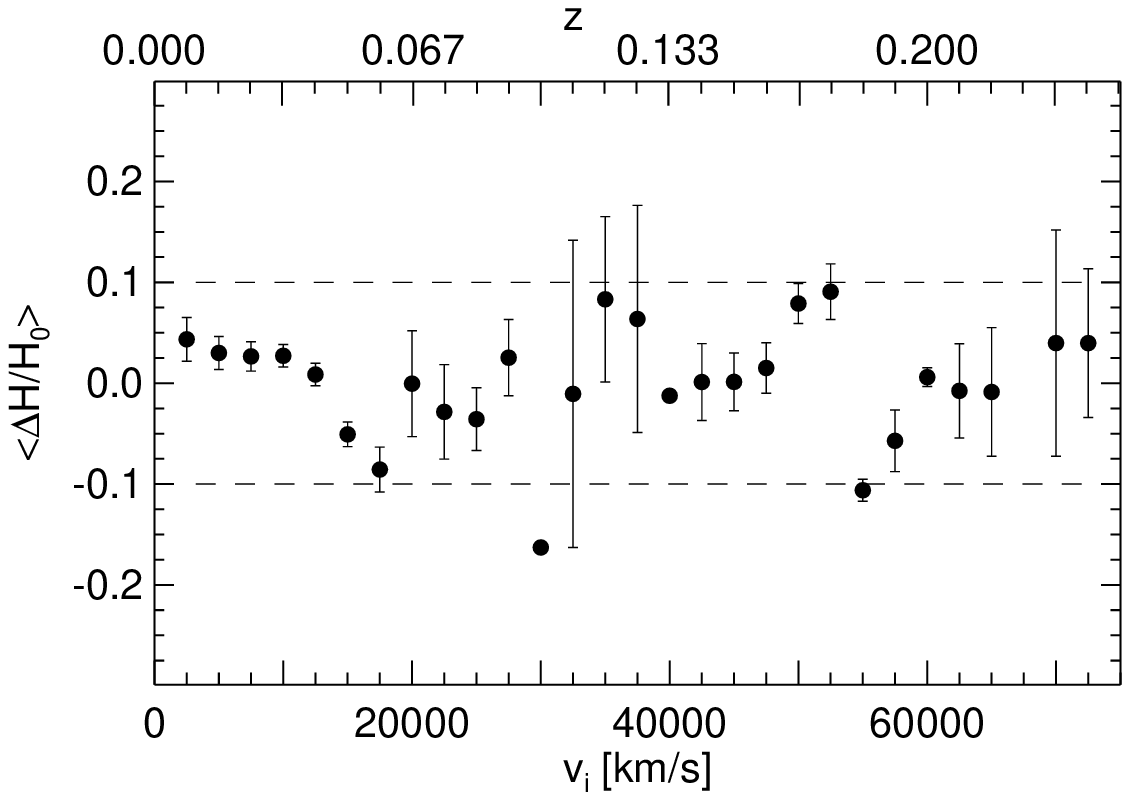,width=\floatwidth}}
\caption{The variation of $H_0$ with redshift derived from
relative distances. a) Out to $35\,000\kms$ using relative
cluster distances, SNe\,Ia, and first-ranked cluster galaxies.
b) Out to $72\,500\kms$; beyond $35\,000\kms$ the data depend
only on first-ranked cluster galaxies.}\label{fig:1}
\end{figure}

Inspection of Fig.~\ref{fig:1} strongly suggests that the
Hubble constant decreases from $1000\kms$ to about $18\,000\kms$
by $\sim7\%$. This trend is independently supported by the
first-ranked cluster galaxies of \citeauthor{Lauer:Postman:94} (1994),
which have not been used here. Beyond $18\,000\kms$ the scatter
becomes large, leaving the possibility of local $\pm10\%$
variations of $H_0$, but the distant overall {\em mean\/} of $H_0$ lies
close to the value found at $10\,000\kms$.

   The consequences for the calibration of $H_0$ are clear.
Full-sky samples with $v>1000\kms$ yield the cosmic value of
$H_0$ to within $\pm5\%$. The best mean cosmic value is found
near $10\,000\kms$ or from sufficiently large samples beyond
$20\,000\kms$.

\section{The Calibration of \boldmath$H_0$}
\label{sec:3}
The interlaced construction of the distance scale out to
$30\,000\kms$ and beyond is schematically shown in Fig.~2.
Brief comments on the individual steps, labelled \point{1} to
\point{9}, proceed as follows.

\noindent \point{1}
The distance of LMC is the first fundamental step outward. The best
values come from the purely geometrical distance of the ring of
SN\,1987A \cite{Panagia:etal:96}, from Cepheids in various passbands,
and RR\,Lyr stars. The zeropoint of the Cepheid P-L relation is
based on Cepheids in Ga\-lac\-tic clusters \cite{Sandage:Tammann:71,Feast:95},
on stellar radii \cite{DiBenedetto:97}, on the Baade-Becker-Wesselink
method \cite{Laney:Stobie:92} and on Hipparcos trigonometric
parallaxes \cite{Madore:Freedman:98,Sandage:Tammann:98}.
The zeropoint of the RR Lyr star P-L relation in function of
metallicity comes from physical considerations \cite{Sandage:93},
the Baade-Becker-Wesselink method \cite{Sandage:Cacciari:90},
and from globular clusters fitted to the Hipparcos-calibrated
main sequence of subdwarfs
\cite{Reid:98,Gratton:etal:97,Pont:etal:98}.
Individual LMC moduli are compiled in \citeauthor{Federspiel:etal:98}
(1998) and give $(m-M)=18.54\pm0.03$.
This value is also in good agreement with the position of the
red-giant tip \cite{Lee:etal:93,Tammann:96}.
Conservatively $(m-M)=18.50$ has been adopted in the following. This
value is secure to within $\pm5\%$ and is uncontroversial.

\def\floatwidth{0.72\textwidth}
\begin{figure}[t]
\centerline{\psfig{bbllx=11pt,bblly=23pt,bburx=530pt,bbury=773pt,%
file=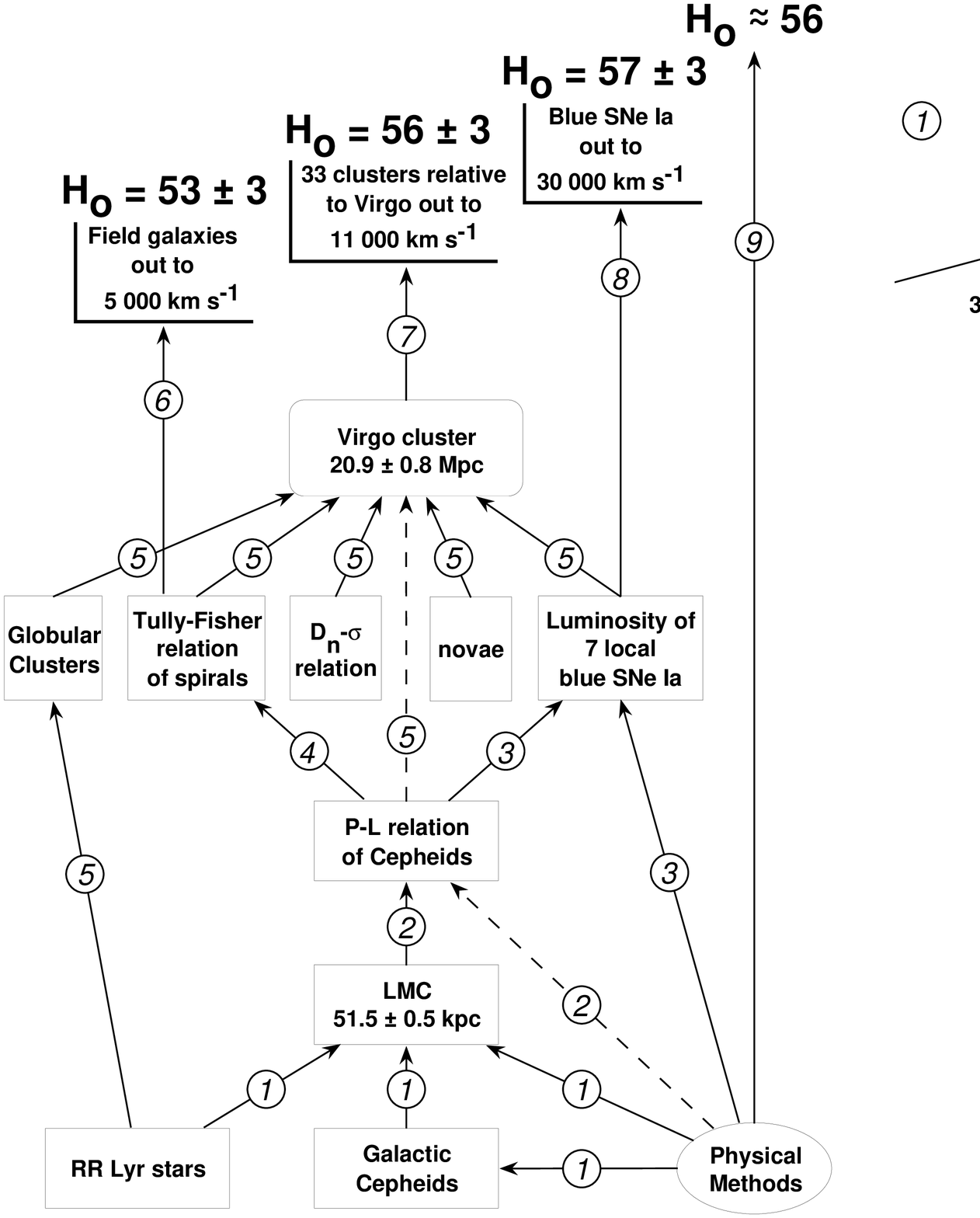,width=\floatwidth,clip=}}
\caption{Schematic presentation of the interlaced extragalactic
  distance scale}\label{fig:2}
\end{figure}

\noindent \point{2}
The distance of LMC combined with the good photometry of many of its
Cepheids yields the calibrated P-L relation at different wavelengths
\cite{Sandage:Tammann:71}. For the $HST$ observations in $I$ and $V$
the P-L relations of \citeauthor{Madore:Freedman:91} (1991) are
generally used; this has the advantage that any divergence in the
derived value of $H_0$ cannot be blamed on the use of different
P-L relation.

   Typical errors of individual Cepheid distances derived from $HST$
data are $\pm10\%$ due to the width of the instability strip and
restricted sample size and due to absorption. For the luminosity
calibration of SNe\,Ia (cf. \point{8}) the errors are smaller because
only apparent distance moduli are used. --- Attempts to improve
Cepheid distances with the help of a period-luminosity-color (PLC)
relation (e.\,g. \citeauthor{Kochanek:97}, 1997) are doomed because
stellar evolution models combined with a pulsation code show that the
basic assumptions going into a PLC relation are not met
\cite{Saio:Gautschy:98}. The same models show that the much-discussed
metallicity has minimal effect on the P-L$_{\rm bol}$ relation;
remaining metallicity effects enter only through the bolometric
correction.

\noindent \point{3}
The Cepheid distances from $HST$, most relevant for $H_0$, concern the
six nearby galaxies which have produced seven well studied blue
(i.\,e. ``Branch - normal'') SNe\,Ia (cf. \point{8}).
Their mean luminosity is \cite{Saha:etal:97,Saha:98}
$M_{\rm B}(\max)=-19.54, M_{\rm V}(\max)=-19.50$ with an impressively
small scatter of $0\mag16$ which -- in addition to the arguments
presented in \point{8} -- demonstrates that blue SNe\,Ia are highly
useful standard candles.

   The luminosity of SNe\,Ia can also be determined from physical
explosion models. The state-of-the-art light curve fitting method by
\citeauthor{Hoeflich:Khokhlov:96} (1996) give $M_{\rm B}(\max)=-19.45$
for the 16 SNe\,Ia to which sufficiently blue models can be fitted. In
a review of all physical determinations of $M(\max)$
\citeauthor{Branch:98} (1998) concludes $M_{\rm B}(\max)\approx
M_{\rm V}(\max)\approx -19.4 - 19.5$.
The close agreement of the astronomical and physical calibrations of
blue SNe\,Ia is most encouraging.

\noindent \point{4}
The calibration of the 21\,cm line width-luminosity (Tully-Fisher, TF)
relation in the B-band can now be based on 18 Cepheid distances most
of which are due to $HST$. They determine the zeropoint of the
relation with a statistical error of only $0\mag05$, but they are
still not enough for the determination of the slope. The latter must
be taken from a {\em complete\/} sample of 49 inclined spirals of the
Virgo cluster \cite{Federspiel:etal:98}.

   In principle redder wavelengths than the B band offer the advantage
of smaller inclination-dependent absorption corrections. However,
this advantage is entirely offset by the steeper slope of the TF relation
at longer wavelengths \cite{Schroeder:96}. Moreover the restricted
number of available magnitudes at longer wavelengths and their possible
inhomogeneity, particularly in the $I$ and $H$ bands, preclude the
all-important construction of {\em complete\/} samples.
Therefore absolute TF distances of the Virgo cluster and of field
galaxies are problematic if derived from magnitudes other than $B$.

\noindent \point{5}
The distance of the Virgo cluster can be derived, as shown in
Fig.~\ref{fig:2}, in six different ways:

\noindent a)
Cepheid distances give so far only a rather poorly determined cluster
distance. The reason is the important depth effect of the
cluster. Four of the five galaxies with distances from Cepheids have
been selected because they are well or even extremely well (NGC\,4571)
resolved and {\em expected\/} to lie on the near side of the
cluster. Their mean distance is $16.1\pm0.4\;$Mpc \cite{Freedman:etal:98}.
One galaxy, NGC\,4639, has been selected irrespective of resolution
and its distance is $25.1\pm2.5\;$Mpc \cite{Saha:etal:97}.
The position of the four resolved galaxies on the near side and of the
poorly resolved NGC\,4639 on the far side of the cluster is
independently confirmed by their relative TF distances
\cite{Federspiel:etal:98}. One can therefore infer only that the
{\em center\/} of the cluster lies roughly at $(m-M)=31.5\pm0.5$
($20\;$Mpc). A preliminary way out is to use the mean distance of the
Leo group of $(m-M)=30.27\pm0.12$, based now on three galaxies with
Cepheid distances, and to step up this value by the modulus difference
of $\Delta(m-M)=1.25\pm0.13$ \cite{Tammann:Federspiel:97} between the
Leo Group and the Virgo cluster, giving $(m-M)_{\rm Virgo}=31.52\pm0.21$.

\noindent b)
Eight SNe\,Ia with known maximum magnitudes are
available. Three of these have occurred in E/S0 galaxies; they are
known to be underluminous by $0\mag18$. If the
latter are adjusted to SNe\,Ia in spirals to conform with the
calibrating SNe\,Ia, the mean apparent magnitude of the eight SNe\,Ia
becomes $<\!m_{B}(\max)\!> =11.91\pm0.16$ and $<\!m_{V}(\max)\!>
=11.84\pm0.17$.
With the calibration in \point{3} the mean cluster modulus becomes then
$31.39\pm0.17$.

\noindent c)
The TF relation in $B$ magnitudes of a {\em complete\/} sample of 49
cluster galaxies, combined with the TF calibration in \point{4}, gives
$(m-M)_{\rm Virgo}=31.58\pm0.24$ \cite{Federspiel:etal:98}. The
relatively large error allows for a number of systematic effects like
sample selection, propagation of observational errors, and inclination
and color differences between calibrators and cluster galaxies.

\noindent d\,-\,f)
Other useful determinations of the Virgo cluster distance come from
the luminosity function of globular clusters, the D$_{\rm n}-\sigma$
relation of the bulges of S0 and spiral galaxies, and from
novae. Because of the restricted space here the reader is referred to
\citeauthor{Sandage:Tammann:97} (1997) and
\citeauthor{Tammann:Federspiel:97} (1997). The results of the six
methods a)\,--\,f) are compiled in Table~\ref{tab:1}.
%
\begin{table}[t]
\caption{The Virgo cluster modulus from various methods}
\label{tab:1}
\begin{center}
\begin{tabular}{lll}
\hline
        Method               & $(m-M)_{\rm Virgo}$ & Hubble type \\
\hline
        Cepheids              &  $31.52 \pm  0.21$ &     S       \\
        SNe\,Ia               &  $31.39 \pm  0.17$ &     E, S    \\
        Tully-Fisher          &  $31.58 \pm  0.24$ &     S       \\
        Globular Clusters     &  $31.67 \pm  0.15$ &     E       \\
        D$_{\rm n} - \sigma$  &  $31.85 \pm  0.19$ &     S0, S   \\
        Novae                 &  $31.46 \pm  0.40$ &     E       \\
\hline
        Mean:                 &  $31.60 \pm  0.08$ &
 ($\Rightarrow 20.9\pm0.8\;$Mpc) \\
\hline
\end{tabular}
\end{center}
\end{table}

   The surface brightness fluctuation (SBF) method is sometimes
advertised as a distance indicator. However, its reliability has not
been sufficiently demonstrated yet for distances beyond $10\;$Mpc. The
SBF distance of NGC\,7331 \cite{Tonry:97} is $0\mag55$ smaller than
its Cepheid distance \cite{Hughes:97}, and the proposed Virgo cluster
modulus \cite{Tonry:97} is smaller than that in Table~\ref{tab:1} by
the same amount. The applicability of the method will still depend on
a large sample of Virgo cluster members to decide if the same relation
applies for ellipticals and bulges of spiral galaxies. If not, the
zeropoint calibration of the method will depend on an adopted Virgo
cluster distance.

   It has also been suggested that the luminosity function of the
shells of planetary nebulae (PNe) in the light of the $\lambda\;
5004\;$\AA{} line had a magic cutoff luminosity which could be used as a
universal distance indicator. However, it has been shown that the
available data are at least equally consistent with a roughly
exponential bright tail of the luminosity function such that the
brightness of the brightest PNe depends on the sample size, i.\,e. on
the luminosity of the galaxy under consideration
\cite{Bottinelli:etal:91,Tammann:93}. This conclusion has been
buttressed by model calculations by \citeauthor{Soffner:etal:96}
(1996). The recent proposal of a Virgo modulus of $(m-M)=30.79\pm0.16$
\cite{Ciardullo:etal:98}, based on a simple cutoff assumption, is
significantly smaller than the Cepheid distance of even the nearest,
highly resolved Virgo spirals (cf. \point{5}a) and is therefore
self-defeating.

\noindent \point{6}
The route to $H_0$ through {\em field galaxies\/} is the most
difficult one. The crux is selection bias. Its origin is the fact that
astronomers work with magnitude-limited catalogs of field galaxies, in
which case the mean luminosity of the catalogued objects increases
with distance. The disastrous consequences of selection (Malmquist)
bias are illustrated, for instance, by a realistic model calculation
by \citeauthor{Hendry:Simmons:90} (1990). They show that denying
the selection bias can lead to $H_{0}=80$, while proper allowance for
bias --- depending on the luminosity scatter of the galaxies ---
yields values of $H_{0}=56$ or even $H_{0}=44$. Two fundamental facts
emerge from this. Neglect of the selection bias always leads to too
high values of $H_0$, and the severity of the error is a strong
function of the luminosity scatter. In the case of cluster distances
one can overcome the problem by working with {\em complete\/} samples
within a given volume (cf.~\point{7}). Blue SNe\,Ia offer the very
important advantage that their luminosity scatter is so small that
$H_0$ will not be significantly overestimated (cf.~\point{8}).

   Analytical corrections for Malmquist bias require knowledge of the
true scatter, which can only be derived from very deep samples, and
must neglect the clumpy distribution of galaxies. Practical ways to
correct for bias are, e.\,g., by \citeauthor{Bottinelli:etal:86a}
\shortcite{Bottinelli:etal:86a,Bottinelli:etal:86b},
\citeauthor{Lynden-Bell:etal:88} (1988),
\citeauthor{Sandage:88} (1988),
\citeauthor{Federspiel:etal:94} (1994),
\citeauthor{Sandage:etal:95} (1995),
\citeauthor{Giovanelli:97a} (1997a), and
\citeauthor{Theureau:etal:97} (1997).
A tutorial on the subject is given by \citeauthor{Sandage:95} (1995).

   Recent determinations of $H_0$ from bias-corrected field galaxies
are compiled in Table~\ref{tab:2}.
%
\begin{table}[t]
\caption{$H_{0}$ from bias corrected field galaxies}
\label{tab:2}
\begin{center}
\begin{tabular}{lll}
\hline
         Method       &   \multicolumn{1}{c}{$H_{0}$}      &    Source \\
\hline
Tully-Fisher                           &   $< 60$    &
   \citeauthor{Sandage:94} 1994   \\
M\,101 look-alike diameters            & $43 \pm 11$ &
   \citeauthor{Sandage:93a} \citeyear{Sandage:93a} \\
M\,31 look-alike diameters             & $45 \pm 12$ &
   \citeauthor{Sandage:93b} \citeyear{Sandage:93b}  \\
Spirals with luminosity classes        & $56 \pm  5$ &
   \citeauthor{Sandage:96a} \citeyear{Sandage:96a}  \\
M\,101, M\,31 look alike luminosities  & $55 \pm  5$ &
   \citeauthor{Sandage:96b} \citeyear{Sandage:96b} \\
Tully-Fisher                           & $55 \pm  5$ &
   \citeauthor{Theureau:etal:97} 1997 \\
Galaxy diameters                       & $50 - 55$   &
   \citeauthor{Goodwin:etal:97} 1997 \\
Tully-Fisher                           & $60 \pm 5$  &
   \citeauthor{Federspiel:98} 1998 \\
\hline
mean                                   & $53 \pm 3$  & \\
\hline
\end{tabular}
\end{center}
\end{table}
%
%
   Field galaxies offer the advantage of full-sky coverage outside the
zone of avoidance. But they are not only the most difficult route to
$H_0$, but also the least satisfactory, having their main thrust as
close as $1000-3000\kms$, i.\,e. at a distance where $H_{0}$\,(local)
may still be a few percent higher than $H_{0}$\,(cosmic)
(cf. Fig.~\ref{fig:1}).

\noindent \point{7}
$H_0$\,(cosmic) from clusters out to $11\,000\kms$. Cluster distances
{\em relative\/} to the Virgo cluster have been compiled from the
literature \cite{Jerjen:Tammann:93}. The sample is increased by
{\em relative\/} $I$-band TF distances of clusters \cite{Giovanelli:97b}.
These relative distances are excellent as seen from their small
scatter about the Hubble line of slope 0.2 in Fig.~\ref{fig:3}.
The scatter is in fact so small that it imposes stringent limits on the
radial component of the peculiar motion of cluster centers
\cite{Jerjen:Tammann:93}.

The best fit to the data in Fig.~\ref{fig:3} is
\begin{equation}\label{equ:1}
   \log cz = 0.2\,[(m-M)_{\rm Cluster} - (m-M)_{\rm Virgo}] + (3.070 \pm 0.024)
\end{equation}
\cite{Federspiel:etal:98}.
From this follows directly
\begin{equation}\label{equ:2}
   \log H_{0}({\rm cosmic}) = -0.2\,(m-M)_{\rm Virgo} + (8.070 \pm 0.024).
\end{equation}
Inserting $(m-M)_{\rm Virgo}=31.60\pm0.08$ yields
\begin{equation}\label{equ:3}
   H_{0}({\rm cosmic}) = 56 \pm 3 \quad \mbox{(internal error)}.
\end{equation}
Note that no use of any velocity of the Virgo cluster has been
made. The value of $H_0$ holds out to $\sim\!10\,000\kms$ where its value is
very close to the large-scale value (Fig.~\ref{fig:1}).
\def\floatwidth{0.67\textwidth}
\begin{figure}[t]
\centerline{\psfig{file=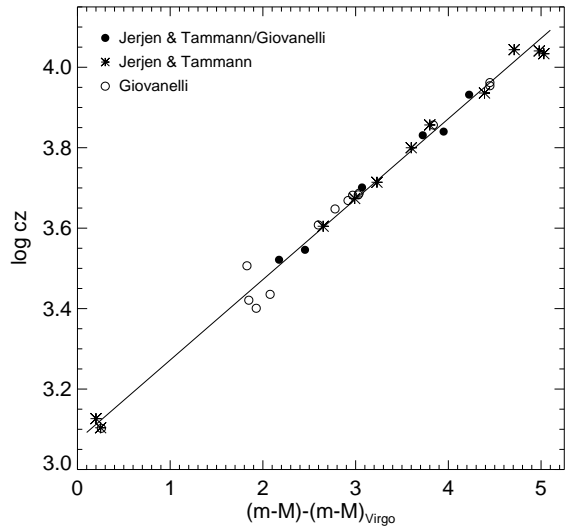,width=\floatwidth}}
\caption{Hubble diagram of 31 clusters with known relative
     distances. Asterisks are data from Jerjen \& Tammann (1993). Open
     circles are from Giovanelli (1997a). Filled circles are the
     average of data from both sources.}\label{fig:3}
\end{figure}

\begin{figure}[t]
\centerline{\psfig{file=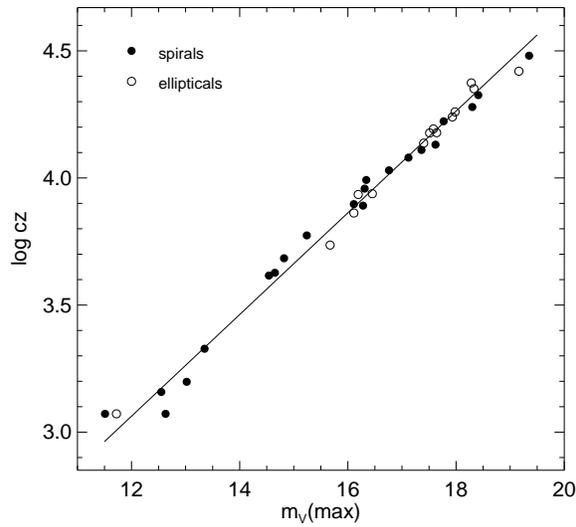,width=\floatwidth}}
\caption{The Hubble diagram of 35 blue SNe\,Ia with photometry after
  1985. The SNe\,Ia in elliptical galaxies are shifted by $0\mag18$.}
  \label{fig:4}
\end{figure}

\noindent \point{8}
$H_{0}$\,(cosmic) from blue SNe\,Ia. The Hubble diagram (in $V$) of 35
blue SNe\,Ia, mainly due to the heroic efforts of the Cerro Tololo
group (Hamuy {\em et al.},\citeyear{Hamuy:etal:96}), is shown in Fig.~\ref{fig:4}.
The small scatter $\sigma(m_{V})$ of $0\mag24$ confirms the conclusion
in \point{3} that blue SNe\,Ia are extremely useful standard
candles. This is also supported by the absence of Malmquist bias,
i.\,e. the Hubble line does not steepen with distance. A fit to the
Hubble diagrams in $B$ and $V$ gives, after simple transformation,
\begin{equation}\label{equ:4}
   \log H_{0} = 0.2\,M_{B} + (5.651 \pm 0.011)
\end{equation}
\begin{equation}\label{equ:5}
   \log H_{0} = 0.2\,M_{V} + (5.669 \pm 0.019).
\end{equation}
These equations hold for SNe\,Ia in spirals. Those in E/S0 galaxies
are fainter on average by $0\mag18\pm0\mag05$ in $B$ and $V$ and have
been shifted by this amount. Equations (\ref{equ:4}) and (\ref{equ:5})
can therefore directly be compared with the calibrators in spiral
galaxies from \point{3}. Thus one obtain a mean value from $B$ and $V$
data of
\begin{equation}\label{equ:6}
   H_{0}({\rm cosmic}) = 57 \pm 3.
\end{equation}
Red SNe\,Ia with $(B_{\max}-V_{\max})>0.20$ have been excluded from
the experiment because they are reddened or have peculiar spectra
\cite{Branch:etal:93}. Their unjustified inclusion leads to
second-parameter corrections of the calibration in \point{3}, but
$H_0$ remains in all cases $<65$ (for details see
\citeauthor{Saha:etal:97} 1997; cf. also \ref{sec:4_6}). Within the
claimed accuracy all second-parameter effects of blue SNe\,Ia have
been taken care of by allowing for the underluminosity of $0\mag18$ of
the SNe\,Ia in E/S0 galaxies.

   Of all present methods to derive $H_0$, the route through SNe\,Ia
   deserves the highest weight.

\noindent \point{9}
$H_0$\,(cosmic) from {\em Physical Methods}. One distinguishes between
astronomical and physical distance determinations. The former depend
always on some adopted distance of a celestial body, be it only the
Astronomical Unit in the case of trigonometric parallaxes. Physical
methods derive the distance solely from the observed physical or
geometrical properties of a specific object.

   In the foregoing reference has been made to physical luminosity and
distance determinations of the SN\,1987A remnant, Cepheids, RR\,Lyr
stars, and SNe\,Ia. But in addition there are a number of physical
distance determinations which lead to the value of $H_0$ over very
large scales. They are still model-dependent, but as the number of
objects increases and the models improve, their weight is steadily
increasing.

   For brevity the most recent physical determinations of $H_0$ are
   compiled in Table~\ref{tab:3}.

%
\begin{table}[ht]
\caption{$H_0$ from Physical Methods}
\label{tab:3}
\begin{center}
\begin{minipage}{0.9\textwidth}
\begin{center}
\begin{tabular}{llc}
\hline
       Method          &   \hspace*{1ex}$H_{0}$   &   Source\footnote{%
        Sources:
    (1) \citeauthor{McHardy:etal:90} 1990;
        \citeauthor{Birkinshaw:Hughes:94} 1994;
        \citeauthor{Lasenby:Hancock:95} 1995
    (2) \citeauthor{Rephaeli:95} 1995;
        \citeauthor{Herbig:etal:95} 1995
    (3) \citeauthor{Holzapfel:etal:97} 1997
    (4) \citeauthor{Lasenby:Jones:97} 1997
    (5) \citeauthor{Myers:etal:97} 1997
    (6) \citeauthor{Rephaeli:Yankovitch:97} 1997
    (7) \citeauthor{Falco:etal:97} 1997
    (8) \citeauthor{Nair:96} 1996
    (9) \citeauthor{Keeton:Kochanek:97} 1997
   (10) \citeauthor{Kundic:etal:97} 1997
   (11) \citeauthor{Schechter:etal:97} 1997
   (12) \citeauthor{Lineweaver:98} 1998
   (13) \citeauthor{Webster:etal:98} 1998
}
                                                             \\
\hline
  Sunyaev-Zeldovich effect                    &             &      \\
         \hspace*{1cm} for cluster A 2218     & $45 \pm 20$ &  (1) \\
         \hspace*{1cm} for 6 other clusters   & $60 \pm 15$ &  (2) \\
         \hspace*{1cm} cluster A 2163  &  $78\,(+54,\,-28)$ &  (3) \\
         \hspace*{1cm} 2 clusters             & $42 \pm 10$ &  (4) \\
         \hspace*{1cm} 3 clusters             & $54 \pm 14$ &  (5) \\
         \hspace*{1cm} incl. relativ. effects & $44 \pm  7$ &  (6) \\
\noalign{\smallskip}
  Gravitational lenses                        &             &      \\
         \hspace*{1cm} QSO 0957 + 561         & $62 \pm 7$  & (7) \\
         \hspace*{1cm} B 0218 + 357           & $52-82$     & (8) \\
         \hspace*{1cm} PG 1115 + 080          & $60 \pm 17$ & (9) \\
         \hspace*{1.9cm} ''                   & $52 \pm 14$ & (10) \\
         \hspace*{1.9cm} ''                   & $62 \pm 20$ & (11) \\
\noalign{\smallskip}
  MWB fluctuation spectrum                    & $58 \pm 11$ & (12) \\
         \hspace*{1.9cm} ''                   & $47 \pm  6$ & (13) \\
\hline
\end{tabular}
\end{center}
\end{minipage}\end{center}
\end{table}

Following \citeauthor{Rephaeli:Yankovitch:97}~(1997) all previous
values of $H_0$ from the SZ effect should be lowered by $\sim\! 10$
units due to relativistic effects.

   The overall impression from the values in Table~\ref{tab:3} is that
   $H_0$ will settle around $H_0\approx56$.

\section{\boldmath$H_0=73$?}
\label{sec:4}
%
\citeauthor{Freedman:etal:97} (1997) have proposed seven arguments
why $H_0$ should be 73 (cf. also \citeauthor{Freedman:etal:98}
1998). These arguments are briefly discussed in the following.

\subsection{$H_{0}=80\pm17$ from the Virgo cluster.}
\label{sec:4_1}
This high value can only be suggested by combining a low cluster
distance of $17.8\;$Mpc, relying heavily on Cepheids of near-side
cluster members, and an outdated mean cluster velocity of
$1404\kms$. The best cluster velocity, corrected for all local
effects, follows from equation~(\ref{equ:1}) by setting $(m-M)_{\rm
Cluster} - (m-M)_{\rm Virgo}=0$, i.\,e. $1175\pm85\kms$. This with
$17.8\;$Mpc gives $H_{0}=66$ and with a more realistic cluster
distance from Table~\ref{tab:1} $H_{0}=56\pm4$.

\subsection{$H_{0}=77\pm16$ from Coma via Virgo.}
\label{sec:4_2}
The value depends, of course, on the adopted distance of the Virgo
cluster. With a Virgo distance of $20.9\pm0.8$ (Table~\ref{tab:1}) one
obtains $H_{0}=66\pm10$ with a large error due to the peculiar motion
of the Coma cluster and to some extent also due to the error of the
adopted relative distance Coma-Virgo.

\subsection{$H_{0}=72\pm18$ from the Fornax cluster.}
\label{sec:4_3}
The result comes from equating the single Cepheid distance of the
spiral galaxy NGC\,1365 of $18.4\pm1.0\;$Mpc \cite{Silbermann:etal:98}
with the {\em mean\/} cluster distance. There is no basis for this ad
hoc assumption.
A compilation of 30 distance determinations of the Fornax cluster over
the last 20 years actually indicates that the E/S0 members are
$(1.17\pm0.05)$ times more distant than the spiral galaxies
\cite{Tammann:Federspiel:97}. Taking this at face value one finds
$21.5\pm1.5\;$Mpc (cf. also \ref{sec:4_6}) for the E/S0 galaxies and
hence for the cluster center. This reduces the proposed value of
$H_{0}=77\pm16$ to $H_{0}=66\pm15$ with a large uncertainty due to the
unknown peculiar motion of the cluster (cf. Section~1).

\subsection{$H_{0}=72\pm8$ from local data.}
\label{sec:4_4}
The solution is dominated by the Virgo cluster (Sec.~\ref{sec:4_1}) and
Fornax clus\-ter (Sec.~\ref{sec:4_3}). The remaining three groups, NGC\,2403,
NGC\,1023, and Leo, lie within $12\;$Mpc and are irrelevant for the
{\em cosmic\/} value of $H_0$.

\subsection{$H_{0}=72\pm8$ from the Jerjen-Tammann clusters.}
\label{sec:4_5}
The cluster sample has since been increased by the relative cluster
distances of \citeauthor{Giovanelli:97a} (1997; cf. Fig.~\ref{fig:3}
and equations (\ref{equ:1}) and (\ref{equ:2})). With an (untenable)
Virgo cluster modulus of $(m-M)=31.25\pm0.20$ from Sec.~\ref{sec:4_1},
equation (\ref{equ:2}) yields $H_{0}=66\pm7$ (not 72!), but even this
is too high because the inserted Virgo modulus is too small.

\subsection{$H_{0}=67\pm8$ from SNe\,Ia.}
\label{sec:4_6}
The authors follow in principle route \point{3}\,$+$\,\point{8}, but
they reject two calibrating SNe\,Ia on the ground of their
photographic photometry (much of observational astronomy had to be
discarded on that ground), {\em and\/} they speculate that SN\,1980
and SN\,1992A are at the same distance as NGC\,1365. The two SNe\,Ia
have occurred in the Fornax cluster E/S0 galaxies NGC\,1316 and
NGC\,1380, respectively. As argued under \ref{sec:4_3}, there is
independent evidence that the E/S0 galaxies in this cluster are more
distant than the spiral NGC\,1365. But for the sake of the argument
the speculation of a common distance is taken up here. Since blue
SNe\,Ia in early-type galaxies are fainter by $0\mag18$ in $B$ and
$V$ than their counterparts in spirals (cf. \point{8}), the mean
absolute magnitude of SNe 1980 and 1992
becomes $M_{B}=-18.95$ and $M_{V}=-19.00$ {\em if\/} they lied at the
distance of NGC\,1365 and if they had occurred in spirals. This
averaged in with the four remaining calibrators in \point{3} gives
$<\!M_{B}(\max)\!>\, = -19.31\pm0.12$ and
$<\!M_{V}(\max)\!>\, = -19.33\pm0.11$, and inserted into equations
(\ref{equ:4}) and (\ref{equ:5}), which hold for blue SNe\,Ia in {\em
  spirals}, yields a mean value of $H_{0}=62\pm6$.
But even this rather low value is still
internally inconsistent because the luminosity
distribution of the six calibrators used becomes highly non-Gaussian,
defying the basic conclusion of standard candles. It violates also the
models of \citeauthor{Hoeflich:Khokhlov:96} (1996) for blue SNe\,Ia. It is
much more plausible that SNe\,1980N and 1992A have the same standard
luminosity as found in \point{3}, adjusted for early-type parent
galaxies. In that case they require for the Fornax E/S0 galaxies
$(m-M)=31.85$ ($23.4\;$Mpc) in agreement with the independent value
under \ref{sec:4_3}.

\subsection{$H_{0}=73\pm7$ from the TF method.}
\label{sec:4_7}
The authors base their claim on $I$- and $H$-band TF cluster distances
of \citeauthor{Mould:etal:97} (1997) and
\citeauthor{Giovanelli:etal:97} (1997).
These authors consider highly incomplete and hence necessarily biased
cluster samples, which may be useful for {\em relative\/} cluster
distances (cf. also \citeauthor{Kraan-Korteweg:etal:88} 1988).
In fact the {\em relative\/} distances of
\citeauthor{Giovanelli:etal:97}, who include a correction for {\em
  differential\/} bias, are found to be excellent
(cf. Fig.~\ref{fig:3}). However, it is inadmissible to tie these {\em
  biased\/} subpopulations directly to the {\em distance-limited\/}
sample of calibrators with known Cepheids. This necessarily leads to
an underestimate of the cluster distances \cite{Teerikorpi:87}. It has
been demonstrated, for instance, that the 25
sufficiently inclined Virgo galaxies with known
$H$-magnitudes yield a cluster modulus $\sim\!0\mag6$ smaller than the
true value derived from a complete Virgo sample
(\citeauthor{Kraan-Korteweg:etal:88} 1988; their Fig.~4). This
discrepancy, which corresponds to a distance factor of $\sim\!1.3$,
perpetuates then through all cluster distances and immediately gives
$H_{0}\approx 55$. The intermediate step through the Virgo cluster,
i.\,e. from local calibrators to Virgo and from Virgo to more distant
clusters, is necessary because the Virgo cluster is so far the only
cluster for which a large and {\em complete\/} sample of spirals is
available as well as extensive photometry and 21\,cm data.

\subsection{$H_{0}=73\pm7$ from physical models of SNe\,II.}
\label{sec:4_8}
The authors cite the work of \citeauthor{Schmidt:etal:94} (1994; the
also cited paper by R.\,P. Kirshner has not appeared). The result
depends strongly on how the bolometric luminosity is distributed over
the spectrum. The so-called dilation factor is a major stumbling
block. On different assumptions \citeauthor{Baron:etal:95}
\shortcite{Baron:etal:95,Baron:etal:96} have obtained $H_{0}\le
50$. Obviously the method cannot be used at present for a quantitative
discussion of $H_0$ (cf. also \citeauthor{Nadyozhin:98} 1998).

\subsection{$H_{0}=73\pm6$ from the D$_{n}-\sigma$ method.}
\label{sec:4_9}
\citeauthor{Mould:etal:97} (1997) have calibrated the D$_{n}-\sigma$
data of \citeauthor{Faber:etal:89} (1989) using the Leo Group and the
controversial Virgo and Fornax clusters as a zeropoint. Only the Leo
group with three Cepheid distances is secure, but it provides with
only two D$_{n}-\sigma$ distances a shaky basis. If one adopts the
Virgo cluster distance from Table~\ref{tab:1} and $(m-M)=31.85$ as
Fornax cluster modulus (\ref{sec:4_3}), an alternative calibration is
obtained, leading to $H_{0}=63\pm6$. --- It may be noted that the
D$_{n}-\sigma$ method applied to the bulges of S0 and S galaxies
yields a {\em high\/} distance of the Virgo cluster
(Table~\ref{tab:1}).

   A recapitulation of Section~\ref{sec:4_1}-\ref{sec:4_9} gives the
following picture. \ref{sec:4_4} and \ref{sec:4_8} should be
excluded as being too local and too controversial,
respectively. \ref{sec:4_1}, \ref{sec:4_2} and  \ref{sec:4_5}
depend entirely on the adopted small Virgo cluster distance, i.\,e. on
the high weight given to the highly resolved galaxies on the near side
of the cluster (\ref{sec:4_1} depends also on the adopted high Virgo
cluster velocity). \ref{sec:4_3} stands and falls with the assumption
that the Cepheid distance of a single spiral (NGC\,1365) provides a
useful {\em mean\/} distance of the E/S0 galaxies of the Fornax
cluster. The same assumption affects \ref{sec:4_6} by about 15\%. The
small Virgo {\em and\/} Fornax distances are essential for the high
value of $H_0$ in \ref{sec:4_9}. The remaining Section~\ref{sec:4_7}
is a textbook illustration of Malmquist bias.

\section{Conclusions}
\label{sec:5}
%
A Test for the variation of $H_0$ with distance suggests a decrease by
$\sim\!7\%$ from $1000 < v \le 18\,000\kms$. At
$v=10\,000\kms$ $H_0$ goes through a value close to the mean over very
large scales.

   A system of three interconnected distance scales (field galaxies,
cluster distances relative to the Virgo cluster, and most
significantly blue SNe\,Ia) give $H_{0}$ (cosmic) $=57\pm7$
(external error). Physical distance determinations from the SZ effect,
gravitationally lensed quasars, and MWB fluctuations scatter about the
same value.

   A discussion of proposed high values of $H_0$ shows that
disagreement focuses on two topics: 1) the true distance of the Virgo
cluster, and 2) the appreciation of the Malmquist bias. One may add as
item 3) the distance of the E/S0 galaxies in the Fornax cluster; the
latter has lower priority because the peculiar motion of this cluster
is unknown, and it is poorly tied into the relative distance scale of
other clusters.

\bigskip
\noindent {\small {\bf Acknowledgement:}
Financial support of the Swiss National Science Foundation
is gratefully acknowledged. The author thanks his colleagues
in the $HST$ team for the luminosity calibration of SNe\,Ia,
i.\,e. Dres. A.~Sandage, A.~Saha, L.~Labhardt F.\,D.~Macchetto,
and N. Panagia, as well as the many collaborators behind the scenes at
the STScI; much of the present understanding of $H_0$ depends on their
work. He also thanks Mr.~Bernd Reindl for his excellent help in all
computational and technical matters.}

%
%

\end{document}